# Controlled Growth of Large-Area Bilayer Tungsten Diselenides with Lateral *P-N* Junctions

*Srinivas V. Mandyam,*[⊥,†] *Meng-Qiang Zhao,*[⊥,†] *Paul Masih Das,*[†] *Qicheng Zhang,*[†] *Christopher C. Price,*[§] *Zhaoli Gao,*[†] *Vivek B. Shenoy,*[§] *Marija Drndić,*[†] *Alan T. Charlie Johnson*[†,*]

[†] Department of Physics and Astronomy, University of Pennsylvania, 209 South 33rd Street, Philadelphia, PA 19104, USA

[§] Department of Materials Science and Engineering, University of Pennsylvania, 3231 Walnut St., Philadelphia, PA 19104, USA

AUTHOR INFORMATION

[⊥] These authors contributed equally to this work.

**Corresponding Author**

*Email: cjohnson@physics.upenn.edu

ABSTRACT: Bilayer two-dimensional (2D) van der Waals (vdW) materials are attracting increasing attention due to their predicted high quality electronic and optical properties. Here we demonstrate dense, selective growth of WSe$_2$ bilayer flakes by chemical vapor deposition with the use of a 1:10 molar mixture of sodium cholate and sodium chloride as the growth promoter to control the local diffusion of W-containing species. A large fraction of the bilayer WSe$_2$ flakes showed a 0 (AB) and 60º (AA') twist between the two layers, while moiré 15 and 30º-twist angles were also observed. Well-defined monolayer-bilayer junctions were formed in the as-grown bilayer WSe$_2$ flakes, and these interfaces exhibited *p-n* diode rectification and an ambipolar transport characteristic. This work provides an efficient method for the layer-controlled growth of 2D materials, in particular, 2D transition metal dichalcogenides and promotes their applications in next-generation electronic and optoelectronic devices.

Two-dimensional (2D) van der Waals (vdW) materials fabricated by stacking of homogeneous or heterogeneous monolayer "building blocks" are attracting increasing attention, since their physical and chemical properties offer promise for use in atomically thin electronic and optoelectronic devices.[1-9] For example, vertical stacking of 2D $MoS_2$ and $WSe_2$ flakes leads to the formation of atomically thin *p-n* heterojunctions.[10, 11] A stack of two graphene layers at a 'magic' twist angle exhibits flat electronic energy bands near the Fermi energy, resulting in correlated insulating states at half-filling and unconventional superconductivity upon electrostatic doping of the material away from the correlated insulating state. [12, 13]

For stacks of homogeneous 2D vdW materials, the number of layers plays a crucial role in the ultimate optoelectronic properties.[1, 14-17] Since the change in properties from monolayer to bilayer is typically more significant than that resulting from additional layers, precise control over bilayer formation is of great interest.[18-21] For example, for many 2D transition metal dichalcogenides (TMDs), the bandgap is direct for the monolayer and indirect for the bilayer.[16, 22, 23] Also bilayer graphene has a widely tunable bandgap unlike the monolayer.[24] Note that the band offsets between different TMD materials are significant, which could inhibit carrier transport.[25] In contrast, homogeneous bilayer TMDs are expected to exhibit superior performance for applications in optoelectronic devices such as light-emitting diodes, laser diodes, and solar cells.[18, 26-28]

A variety of approaches have been developed to create bilayer TMDs, including mechanical exfoliation from the bulk,[29] mechanical transfer,[21, 30] drop casting[31] or layer-by-layer assembly[32] using exfoliated suspensions, and chemical vapor deposition (CVD).[1, 14, 18] CVD is of particular interest because it not only provides high-quality and scalable TMDs but also can lead to cleaner interfaces between the layers and more intimate electronic coupling. The TMD layer number is affected by many CVD growth parameters, such as substrate surface chemistry,[33-34] partial pressure of the precursors,[35, 36] and growth temperature.[18, 37] For example, layer-controlled CVD growth of large-area $MoS_2$ films (including bilayer) was achieved by precise control of the $MoCl_5$ precursor[35] and by treating the $SiO_2$ substrate surface with an oxygen plasma.[33] CVD growth of TMD flakes with 1 – 4 layers has been achieved through the choice of growth temperature.[18, 37] In contrast to large-area continuous films, TMD flakes can contain both monolayer (1L) and bilayer (2L) regions, and the 1L-2L junction exhibits appealing optoelectronic properties for further device applications.[26-28] So far, reported methods for synthesis of bilayer TMD flakes have resulted in a mixture of monolayer, bilayer, and few-layer flakes, with no more than 80% bilayer content.[18, 37, 38] Obtaining efficient growth of high-density homogeneous bilayer TMD flakes with well-defined 1L-2L junctions remains a challenge.

The complex growth environment during CVD complicates precise control over the final TMD structure. Recently, a multiscale model for vertical growth of 2D vdW materials was proposed that indicated that temperature and vapor adatom flux are the primary parameters affecting the final TMD structure, in agreement with experiments.[20] Among the TMDs, 2D $WSe_2$ is of particular interest because of its tunable transport behaviors and suitability for a variety of electronic and optoelectronic applications.[37, 39] Here, we report an approach for large-area growth of high-density, almost purely bilayer WSe$_2$ flakes using a mixture of sodium cholate and sodium chloride as the growth promoter. These bilayer WSe$_2$ flakes provide abundant lateral 1L-2L junctions which exhibit promising *p-n* diode-like rectification behavior, suggesting their potential use in atomically thin electronics and optoelectronics.

RESULTS AND DISCUSSION

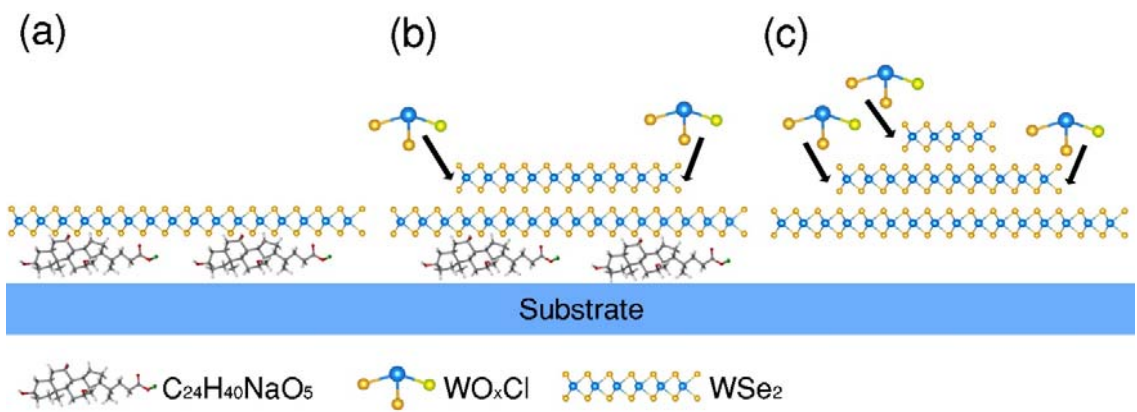

**Figure 1** Schematic showing how the choice of promoter affects CVD growth of WSe$_2$: (a) Sodium cholate promoter leads to growth of monolayer WSe$_2$; (b) The correct mixture of sodium cholate and NaCl leads to the formation of volatile oxyhalides (WO$_x$Cl) and enhanced growth of bilayer WSe$_2$; (c) Pure NaCl promoter leads to an overabundance of oxyhalides and growth of multilayer WSe$_2$.

A variety of organic molecules and salts have been explored as growth promoters of TMDs.[40,41] Organic aromatic molecules are expected to enhance the wettability of the growth substrate and lower the free energy for nucleation, which should promote adsorption of transition metal (TM) oxide precursors on the substrate and thus growth of monolayer flakes.[40, 42] In contrast, alkali metal halides can react with TM oxides to form volatile oxyhalides, which promotes the escape of TM precursors into the vapor and facilitates the vertical growth of additional TMD layers.[20, 43, 44] We discovered that a suitable balance of these two competing mechanisms enabled tuning of the layer number in CVD-grown TMD flakes,[20] as illustrated in **Figure 1**. In our growth method, W precursor is pre-deposited on the substrate by spin-coating of an ammonium metatungstate aqueous solution, together with the growth promoters (see details in Methods). When using sodium cholate ($C_{24}H_{40}NaO_5$) as the growth promoter, the aromatic cholate anions enhance the adsorption of $WO_x$ tungsten precursors onto the substrate and deplete W adatom flux in the vapor, leading to preferential growth of monolayer $WSe_2$ (**Figure 1a**).[20, 40, 42] The addition of NaCl transforms some $WO_x$ into oxyhalide ($WO_xCl$), which is more easily evaporated at high temperature, leading to increased W adatom flux from the vapor and enhanced growth of the bilayer (**Figure 1b**).[20, 43, 44] When pure NaCl is used as the growth promoter, a large amount of $WO_xCl$ forms, leading to reduced surface adsorption and further increased W vapor adatom flux, so that multilayer (ML) $WSe_2$ dominates the final structures (**Figure 1c**).

**Figure 2** shows optical micrographs of $WSe_2$ flakes grown using these three different promoter mixtures. The bare substrate and TMD flake regions with different layer numbers in these images are readily identified due to differences in optical contrast.[45] When 25 mM $C_{24}H_{40}NaO_5$ was used as the growth promoter, star-like $WSe_2$ flakes were obtained that were nearly exclusively monolayer (**Figure 2a-b**). Vertical growth of $WSe_2$ was induced by adding NaCl to allow for enhanced W diffusion from the vapor. When the promoter solution contained 25 mM sodium cholate and 125 mM NaCl, the growth product evolved into monolayer WSe$_2$ with small bilayer regions (**Figure S1a**). When the promoter solution contained 25 mM C$_{24}$H$_{40}$NaO$_5$ and 250 mM NaCl, the growth was optimized and yielded predominantly bilayer WSe$_2$ with sharp 1L-2L boundaries (**Figure 2c-d**).

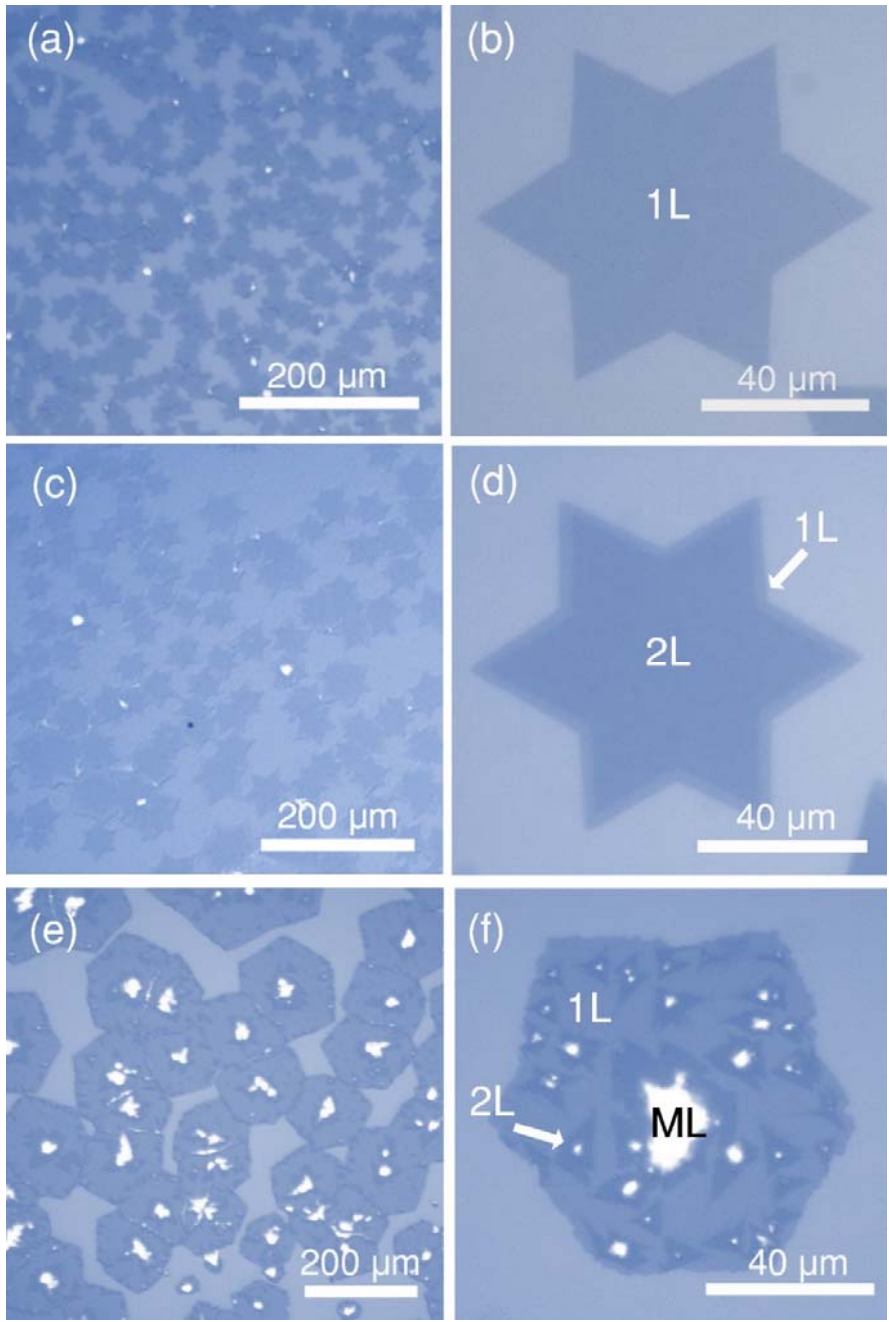

**Figure 2** Optical micrographs showing the morphology of as-grown WSe$_2$ flakes. (a,b) Use of pure sodium cholate as promoter leads to the formation of monolayer flakes. (c,d) A 1:10 mixture of sodium cholate and NaCl as promoter leads to a high coverage of bilayer flakes. (e,f) Use of pure NaCl as the promoter leads to predominantly multi-layer flakes.

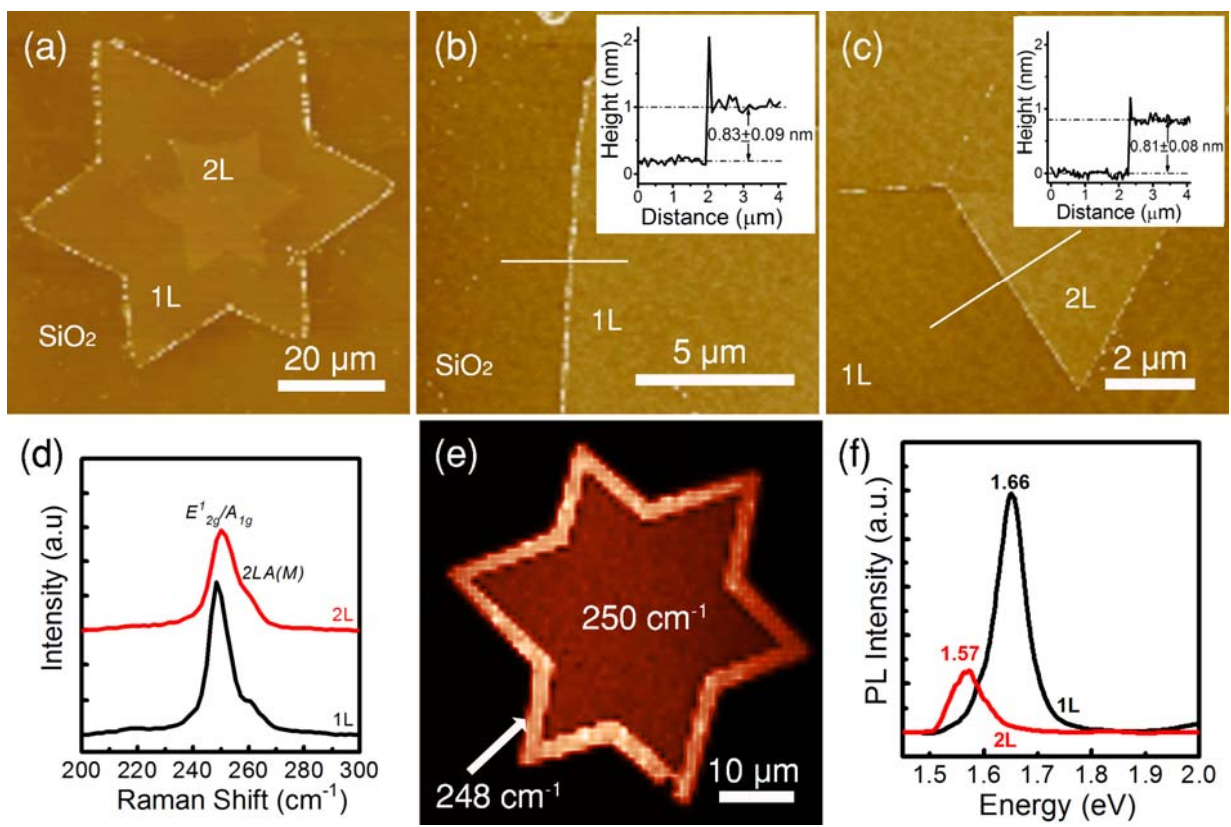

**Figure 3** Atomic Force Microscope images showing (a) a bilayer WSe$_2$ flake on a SiO$_2$ substrate and a zoom in on the (b) substrate-to-1L and (c) 1L-to-2L regions. The measured height values are shown in the corresponding insets. (d) Raman spectra of the 1L and 2L regions of bilayer WSe$_2$; (e) Combined Raman intensity map at 248 and 250 cm$^{-1}$ for a bilayer WSe$_2$ flake; (d) Photoluminescence spectra collected from the 1L and 2L regions of a bilayer WSe$_2$ flake.

After examining >500 flakes over an area of 1×2 cm$^2$, it was confirmed that 100% of the flakes were bilayer WSe$_2$ with a high surface coverage of 67.8 ± 1.5% and a size ranging from 40-80 μm. No monolayer WSe$_2$ flakes were observed, and only a few multilayer regions (bright spots in **Figure 2c**) were found, which was attributed to non-uniform distribution of W precursor. The formation of six-point star-like morphology of bilayer WSe$_2$ was ascribed to a higher Se:W ratio during the growth compared to the level that leads to conventional triangular or hexagonal morphologies.[46, 47] These star-like flakes are composed of six diamond-shaped single crystals with mirror grain boundaries (GBs) between each other, as confirmed by optical microscopy after partial oxidation to reveal the GBs (**Figure S2**).[46-49] Bilayer WSe$_2$ flakes with other morphologies were also observed, indicating that increasing the local value of the Se:W ratio induced a transformation from triangular to star-like flakes (**Figure S3**).[47] When the NaCl concentration was further increased to 500 mM, a large amount of ML WSe$_2$ regions were observed (**Figure S1b**).

To demonstrate the importance of the C$_{24}$H$_{40}$NaO$_5$ component, we conducted the growth with a 250 mM solution of NaCl as the promoter and observed the formation of multiple bilayer and multilayer regions on large monolayer flakes (**Figure 2e-f**). This result can be explained by greatly enhanced diffusion of W from the vapor due to the formation of oxyhalides and comparatively inefficient surface adsorption due to the absence of cholate ions. We also observed that the flakes were largely hexagonal rather than star-like, indicating a lower Se:W ratio.[47] The ML WSe$_2$ regions became smaller with decreasing NaCl concentration. When the NaCl concentration was reduced to 75 mM, the growth yielded a large amount of randomly distributed 2L regions on large 1L WSe$_2$ flakes (**Figures S4a-b**). At 25 mM of NaCl, the growth consisted nearly exclusively of monolayer star-shape $WSe_2$ flakes decorated with a high density of small 2L regions (**Figures S4c-d**). The growth experiments done with varied NaCl concentrations indicated that the use of pure NaCl as a promoter cannot lead to the growth of high-quality bilayer $WSe_2$ flakes with large, continuous 2L regions. This revealed the essential role of surface adsorption introduced by cholate ions in the controlled growth of bilayer $WSe_2$. Note that all the above mentioned morphologies for the as-grown $WSe_2$ also showed a selectivity of ~100% based on the statistical analysis of >500 flakes over an area of 1×2 $cm^2$. We further applied the bilayer growth method to other TMDs. As shown in **Figures S5**, the use of 25 mM $C_{24}H_{40}NaO_5$ and 250 mM NaCl led to the preferential growth of bilayer $WS_2$ while only monolayer flakes were obtained with 25 mM $C_{24}H_{40}NaO_5$ as the growth promoter.

Atomic force microscopy (AFM) was used to confirm the assignment of monolayer and bilayer $WSe_2$ flakes. A low-magnification AFM image is shown in **Figure 3a**, which reveals distinct 1L and 2L regions. The step heights at the substrate-to-1L and 1L-to-2L interfaces were measured to be 0.83±0.09 and 0.81±0.08 nm, respectively, as expected for bilayer material (**Figure 3b-c**). Raman and photoluminescence (PL) spectroscopy were also used to confirm the assignment. Raman spectra collected from the 1L region of a bilayer $WSe_2$ flake showed peaks at 248 and 260 $cm^{-1}$, which correspond to the $E^1_{2g}/A_{1g}$ and 2LA(M) modes of the monolayer, respectively. For the inner 2L region, these two peaks red-shifted to 250 and 263 $cm^{-1}$ (**Figure 3d**), in good agreement with previous reports.[50] The distribution of 1L and 2L regions was further examined by Raman intensity maps at 248 and 250 cm$^{-1}$. A clear 1L-2L boundary was observed in the combined Raman intensity map shown in **Figure 3e**, which agreed well with the optical image shown in **Figure 2d**. The PL spectrum (**Figure 3f**) acquired from the 1L region of a bilayer WSe$_2$ flake showed a strong peak at an energy of 1.66 eV (wavelength of 749 nm), characteristic of a direct excitonic transition. In contrast, on the 2L region the prominent peak showed a lower intensity, indicative of an indirect bandgap transition, and it was shifted to an excitonic transition energy of 1.57 eV (wavelength of 791 nm).[51]

To investigate the atomic scale structure of bilayer WSe$_2$, scanning transmission electron microscopy (STEM) was employed. STEM high-angle annular dark field (STEM-HAADF) images are widely used to identify detailed atomic arrangement of crystalline structures because its contrast is sensitive to the atomic number and sample thickness.[52] **Figure 4a** shows a low-magnification STEM image of a 1L-2L boundary region of an as-grown WSe$_2$ flake, with an atomically sharp 1L-2L interface (**Figure 4b**). The usual hexagonal arrangement of Se and W atoms, corresponding to the 1H phase, is clearly shown in the 1L region. The 2L region has a 60°-twist angle, as confirmed by the SAED pattern (**Figure 4c**) and fast Fourier transform (FFT) decomposition (**Figures S6a-c**), with the resulting atomic model shown in **Figure 4d**. Disordered hexagonal shapes at the 1L-2L boundary are observed, which is ascribed to strain introduced by interlayer vdW interactions.

The atomic structure of a 1L-2L boundary region with a 15°-twist angle is shown in **Figure 4e**, and the twist angle is confirmed by SAED (**Figure 4f**) of the 2L region. The 15° twist angle leads to the formation of moiré patterns, as seen in the STEM-HAADF image and the atomic model of **Figure 4g**. Examples of AB stacking (0º twist) were also observed (**Figures S6g-i**). As mentioned

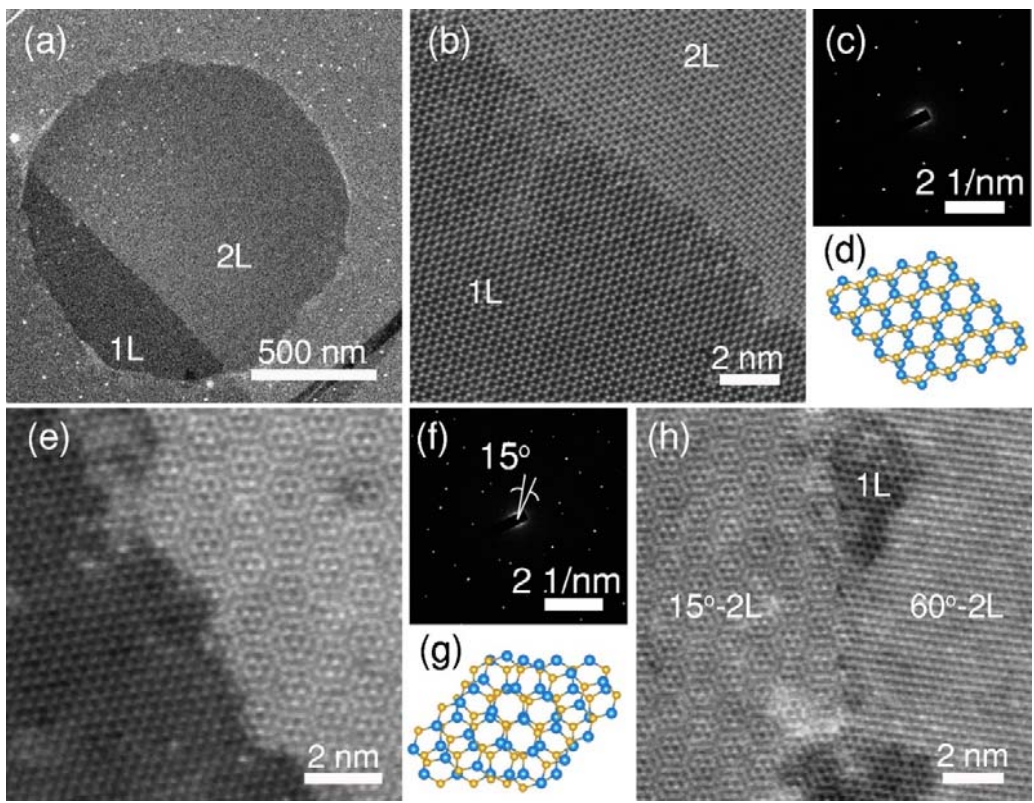

**Figure 4** (a) Low-magnification scanning transmission electron microscope (STEM) image showing a 1L-2L junction in a CVD-grown bilayer $WSe_2$ flake; (b) STEM high-angle annular dark field (STEM-HAADF) image showing a 1L-2L boundary with a 60º (AA')-twist angle in the 2L region with (c) corresponding SAED pattern and (d) an atomic model; (e) STEM-HAADF image showing a 1L-2L boundary with a 15º-twist angle in the 2L region with (f) corresponding SAED pattern and (g) an atomic model; (h) STEM-HAADF image showing the 2L region with a GB of a star-like bilayer $WSe_2$ flake, in which 15 and 60º-twist angles are observed at the two sides of the GB, respectively.

above, star-like WSe$_2$ flakes are composed of six diamond-like single crystals separated by GBs. The atomic structure of one such GB is shown in **Figure 4h**, where neighboring 2L regions exhibit twist angles of 15 and 60º. Although part of the GB was damaged during transfer, which exposed the bottom 1L region, seamless connection between the top neighboring WSe$_2$ layers is still observed. This STEM image clearly reveals that different twist angles exist in a single as-grown bilayer WSe$_2$ flake. The stacking orientation of two TMD layers can be used to manipulate their electrical, optical, and vibrational properties.[21, 53]

We conducted a SAED survey of ~ 100 bilayer flakes and found that only twist angles of 0°/60° (these twist angles cannot be differentiated with SAED alone), 15°, and 30° were represented (**Figure S7**). A twist angle of 0°/60° could indicate any of three distinct stacking structures: AA (0°; known to be thermodynamically unstable[54-56]), AB (0°), and AA' (60°). The strong majority (~ 84%) of flakes showed a twist angle of 0°/60°, with smaller fractions having a twist of 30º (12%) or 15º (4%). A summary of the twist angles and corresponding atomic models is provided in **Figure S8**. Due to the complex growth environment, a good control over the twist angles between TMD layers by CVD is still a challenge.

We examined the energetics of different bilayer orientations using our previously developed thermodynamic criterion for bilayer CVD growth of TMDs:[20] $\varepsilon_{L1:S} - \varepsilon_{L1:L2} > \eta(\gamma/L2 - \gamma/L1)$. Here, $\varepsilon_{L1:L2}$ is the TMD stacking energy, $\varepsilon_{L1:S}$ is the TMD/substrate interaction energy, $\gamma$ is the WSe$_2$ Se-terminated zig-zag edge energy, $\eta = 4/3^{1/2}$ is a geometric factor for diamond flakes, and L1, L2 are the edge lengths of the bottom and top layers, respectively. This criterion determines whether additional growth of the top layer is thermodynamically favorable based on the sizes of the bottom and top layers (constrained such that L2 < L1), and the energetics of edge formation and stacking of the TMD layers. We chose diamond flakes to match the experimental observations, and we calculated the Na-doped Se-terminated zigzag edge energy to be 0.57 eV/ Å. **Figure S9** shows a plot of the criteria over the values of L1 and L2 for the different rotational alignments of the WSe$_2$ bilayer observed during growth. We see that the minimum size for a stable bilayer nucleus to grow is ~1.5 nm, corresponding to a 2D triangular flake containing at least 40 W atoms. Furthermore, critical thermodynamic stability of the top layer (independent of the bottom layer size) is not reached until the top layer flake reaches 4 nm for the AA' and AB stacking orientations, and 6 nm for the 15° and 30º orientation. In between the minimum size and the critical size, the morphology will be kinetically determined, which agrees with the experimental finding that altering the metal diffusion rate through promoter composition determines the preference for bilayer *vs.* monolayer. In particular, increasing the metal diffusivity (through introduction of volatile halide anions) will increase the kinetic probability of bilayer morphology. We also found that the 60º (AA') and 0º (AB) stacking energies are the most favorable and almost identical (**Table 1**), leading to the prediction that they should dominate the population. The next most favorable stackings are the 30 and 15° twisted structures, consistent with their lower observed frequency in the growth. High energy stacking structures, such as 0° (AA), have very large thermodynamic critical sizes (10 nm), consistent with the fact that they are not observed in the experiments.

**Table 1** Calculated vdW stacking energy density for different orientations of WSe$_2$ bilayers, along with lattice constants used.

|  | 0° (AA) | 0° (AB) | 15° (Moiré) | 30° (Moiré) | 60° (AA') |
|---|---|---|---|---|---|
| $\varepsilon_{L1:L2}$ (meV/Å$^2$) | -23.3 | -34.3 | -28.9 | -29.3 | -35.0 |
| $a$ (Å) | 3.29 | 3.29 | 21.54 | 22.87 | 3.29 |

A set of back-gated field effect transistors (FETs), with electrodes 1-5, was fabricated across the 1L-2L domains as well as GBs within a single bilayer flake, as shown in **Figures 5a-b**. The $I_{ds}$-$V_{ds}$ curves of the 1L-2L junction (electrodes 1/2) at different gate voltages ($V_g$) exhibited typical rectification of a *p-n* diode structure (**Figure 5c**), which is ascribed to the different built-in surface potentials of 1L and 2L WSe$_2$ and band structure offsets at their interface.[2, 27] The formation of *p-n* diodes with 1L-2L TMD junctions has been widely reported.[2, 18, 27] In particular, the band structure of WSe$_2$ 1L-2L junctions and the mechanism for the rectification behavior have been documented, and our results are in good agreement with this report.[27] With a gate voltage of -80 V and a forward (reverse) bias of 4 V, the device current is 5.4 µA (- 0.4 µA), yielding a rectification ratio of 13.5. **Figure 5d** presents the $I_{ds}$-$V_g$ transfer characteristic of the same device for different source-drain voltages. The transfer curves exhibit ambipolar behavior, but the conductance in the *p*-branch is about 1000 times larger than in the *n*-branch. This is contributed by both the higher hole mobilities compared to electron for WSe$_2$ and the existence of current transport barrier on the *n*-branch of the *p-n* diode device.[37, 39, 57]

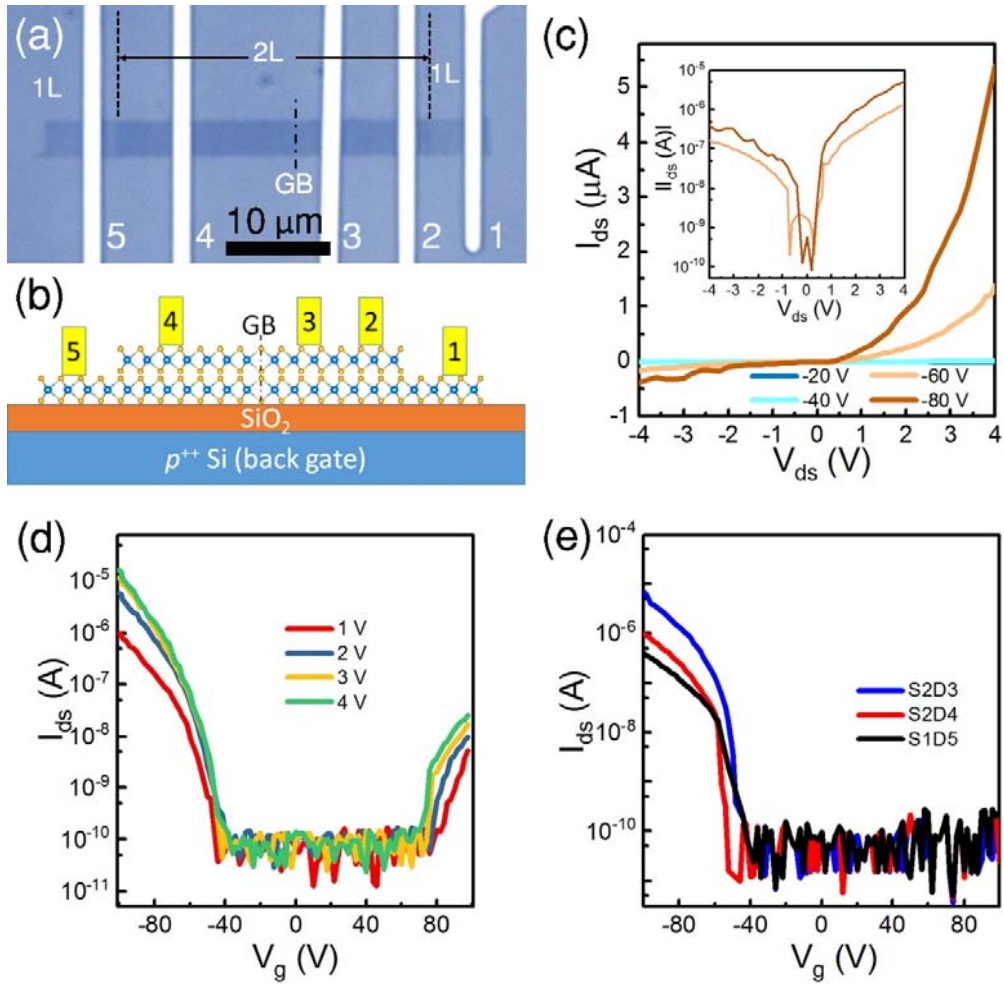

**Figure 5** (a) Optical micrograph and (b) schematic of the fabricated set of FET devices based on as-grown bilayer WSe$_2$; (c) $I_{ds}$-$V_{ds}$ curves at different $V_g$ of the WSe$_2$-based FET with a 1L-2L junction (S1D2); The inset shows the same data on a logarithmic scale. (d) $I_{ds}$-$V_g$ curves at different $V_{ds}$ of the WSe$_2$-based FET with a 1L-2L junction (S1D2); (e) Comparison of the $I_{ds}$-$V_g$ curves for bilayer WSe$_2$-without GBs (S2D3), bilayer WSe$_2$ with a GB (S2D4), and monolayer WSe$_2$ with a GB (S1D5) based FET devices.

We further investigated the electrical transport behavior of monolayer and bilayer WSe$_2$ by measuring devices defined by the electrode pairs 2/3 (S2D3; a bilayer device), 2/4 (S2D4; bilayer device including a GB), and 1/5 (S1D5; monolayer device including a GB). As shown in **Figure 5e**, the $I_{ds}$-$V_g$ curves for all three FET devices exhibited *p*-type behavior. The carrier mobility of FET S2D3 was calculated as 7.9 cm$^2$s$^{-1}$V$^{-1}$, somewhat higher than that for FET S2D4 (5.3 cm$^2$s$^{-1}$V$^{-1}$), reflecting scattering due to the GB between electrodes 2 and 4. FET S2D3 also showed an ON/OFF ratio of ~10$^5$. For the S1D5 FET the carrier mobility was calculated as 3.5 cm$^2$s$^{-1}$V$^{-1}$. An FET device based on monolayer WSe$_2$ without GBs was also fabricated using a bilayer flake with a large monolayer region (**Figure S10a**). This FET exhibited *p*-type behavior with a carrier mobility of 4.3 cm$^2$s$^{-1}$V$^{-1}$ and an ON/OFF ratio of ~5×10$^4$ (**Figure S10b**). The results reflected the lower mobility of monolayer compared to the bilayer [8] and further demonstrated the scattering caused by GBs. The relatively low ON/OFF ratios of the fabricated devices are ascribed to the relatively large OFF-state current (10$^{-10}$–10$^{-11}$ A), which reflects limitations of the instrument, since an OFF current of ~10$^{-12}$ A is frequently observed for similar back-gated WSe$_2$ FETs.[37,57] The $I_{ds}$-$V_{ds}$ curves of the monolayer WSe$_2$ FET at varied gate voltages ($V_g$) confirmed the *p*-type behavior of the material. A smaller current at negative $V_{ds}$ was recorded compared to the positive side, indicating the Schottky contact behavior.[39] However, no significant rectification was observed, consistent with expectations (**Figure S10c**).

CONCLUSION

Controlled large-area CVD growth of high-density bilayer WSe$_2$ flakes was achieved by using an optimized mixture of sodium cholate and sodium chloride as the growth promoter. The best bilayer morphology was obtained when the molar ratio of C$_{24}$H$_{40}$NaO$_5$ to NaCl was 1:10. Further increase in the NaCl concentration was found to lead to the formation of multilayer WSe$_2$ due to the improved diffusion of W precursors. The as-grown WSe$_2$ mainly exhibited a six-point star-like morphology, with well-defined 1L-2L junctions. The majority of the bilayer regions (~ 84%) displayed twist angles of 0° (AB) or 60° (AA'), while moiré 15° and 30°-twist angles were also observed. These CVD-grown 1L-2L WSe$_2$ junctions exhibited a *p-n* diode characteristic with current rectification and an ambipolar transfer characteristic. Our results provide a method for the layer-controlled growth of TMDs, contributing to the fabrication of advanced atomically thin electronic and optoelectronic devices.

METHODS

**Growth of bilayer WSe$_2$.** Bilayer WSe$_2$ flakes were grown directly on a 285 nm SiO$_2$/Si substrate by CVD. First, a mixed aqueous solution of 25 mM sodium cholate, 250 mM NaCl, and 1.5 mM ammonium metatungstate (H$_{26}$N$_6$O$_{40}$W$_{12}$), is spin-coated at 4000 rpm for 60 s onto the SiO$_2$/Si substrate. After that, the substrate was placed in the center of a 1 inch CVD tube furnace and 150 mg of selenium was placed 13 cm upstream from the substrate. Growth occurred at the atmospheric pressure in a flow of 350 sccm of nitrogen gas (99.99% purity) and 15 sccm of H$_2$

(99.999% purity). The furnace temperature was ramped to 900 °C at a rate of 70 °C min$^{-1}$. While the SiO$_2$/Si growth substrate reached 900 °C, the maximum temperature of the selenium was ~220 °C. After a 10 min growth period, the furnace was opened, and the sample was rapidly cooled to room temperature in 700 sccm flowing nitrogen. Growth with other concentrations of promoters, such as 25 mM C$_{24}$H$_{40}$NaO$_5$ + 125 mM NaCl, 25 mM C$_{24}$H$_{40}$NaO$_5$ + 500 mM NaCl, 25 mM NaCl, 75 mM NaCl, and 250 mM NaCl, were conducted in a similar manner, respectively, to investigate the influence of growth promoters.

**Characterization.** AFM analysis was conducted on an atomic force microscope (AFM, Icon Bruker) equipped with a probe with a tip radius of <10 nm (TAP300AI-G, Budgetsensors) to evaluate the height profiles of bilayer WSe$_2$. Raman and PL spectroscopy were performed under ambient conditions using an ND-MDT spectrometer equipped with a laser excitation wavelength of 532 nm. For STEM imaging, bilayer WSe$_2$ flakes were transferred onto a holy-carbon TEM grid using a poly(methyl methacrylate) (PMMA)-assisted transfer process.[58] STEM was performed with a JEOL ARM 200CF equipped with a CEOS corrector (Cs ~ 100 nm) operating at 80 kV to reduce knock-on damage. Images were acquired with a HAADF detector from 54-220 mrad and cleaned using an average background subtraction filter (ABSF) SAED patterns were acquired with a JEOL F200 TEM operating at 200 kV using a selected-area aperture with an effective size at the sample of ~1 μm.

**DFT calculations.** First-principles DFT simulations are carried out using the Vienna *ab initio* simulation package (VASP)[59]. Projector-augmented wave pseudopotentials[60] are used with a cutoff energy of 400 eV for plane-wave expansions. The exchange-correlation is treated using the Perdew-Burke-Ernzerhof (PBE) generalized gradient approximations.[61] The atomistic structures of monolayer and bilayer $WSe_2$ are relaxed using Γ-centered *k*-point meshes of 18x18x1 for primitive cells and 4x4x1 for the twisted Moiré cells. The incommensurate Moiré cells with twist angles of $15^0$ and $30^0$ are strained < 0.4% into commensurate cells to satisfy periodic boundary conditions, and energetic corrections for the introduction of this strain are subsequently added. Long range van der Waals dispersion interactions were treated using the DFT-D3 method developed by Grimme *et. al.*[62, 63] For structural relaxations, the atomic positions of all unit and supercells are optimized until the force components on each atom are less than 0.005 eV/ Å, and the electronic energy is converged within $10^{-8}$ eV. A vacuum spacing of 20 Å was added to prevent interactions between periodic images of the bilayers. Details of edge, substrate, and stacking energy determination can be found in the supporting information of our previous report.[20]

**Device fabrication.** To fabricate FET devices for transport measurements, the as-grown bilayer $WSe_2$ was spin-coated with C4 PMMA (Microchem) under 3000 rpm for 60 s. Electron beam lithography (EBL) and plasma etching were used to define a rectangular region containing the 1L and 2L regions of the bilayer $WSe_2$. Markers in the PMMA were defined by EBL, and then optical microscopy was used to determine the locations of rectangular flake with respect to the markers. A final round of EBL was performed to pattern contacts that were aligned to the 1L and 2L regions, and the contacts were metallized with 5 nm Cr and 40 nm Au deposited by thermal evaporation.

This was followed by lift-off in an acetone bath and a rinse in isopropyl alcohol (IPA). The sample was then dried by a $N_2$ gun and transferred to the measurement system.

**Electrical measurements.** Electrical measurements were performed under ambient condition in a probe station. Current-gate voltage ($I_{ds}$-$V_g$) measurements were carried out using a Keithley 2400 sourcemeter, with a bias voltage of 1, 2, 3, 4 V, respectively. The gate voltage was applied using a Keithley 6487 voltage source. The Current-bias voltage ($I_{ds}$-$V_{ds}$) measurements were carried out using a Keithley 6517A sourcemeter with different gate voltages.

ASSOCIATED CONTENT

The authors declare no competing financial interest.

**Supporting Information**

The Supporting Information is available free of charge on the ACS Publications website at DOI: 10.1021/acsnano.xxxxxxx.

Optical images of the as-grown $WSe_2$ flakes using different growth promoters and partially oxidized bilayer $WSe_2$; Characterizations of the as-grown bilayer $WS_2$ flakes; Analysis of the stacking orientation and atomic structure of bilayer $WSe_2$; SAED patterns of the bilayer $WSe_2$ and corresponding statistical analysis; Atomic models of bilayer $WSe_2$ with varied stacking orientations; Thermodynamic growth diagram for bilayer WSe$_2$ flakes; Characterization data of a WSe$_2$ FET based on the monolayer region.

AUTHOR INFORMATION

**Corresponding Author**

*E-mail: cjohnson@physics.upenn.edu.

**ORCID**

Srinivas V. Mandyam: 0000-0003-1927-7678

Meng-Qiang Zhao: 0000-0002-0547-3144

Paul Masih Das: 0000-0003-2644-2280

Christopher C. Price: 0000-0002-4702-5817

Zhaoli Gao: 0000-0003-3114-2207

Alan T. Charlie Johnson: 0000-0002-5402-1224

**Author Contributions**

⊥ S.M and M.Q.Z contribute equally to this work.

The manuscript was written through contributions of all authors. All authors have given approval to the final version of the manuscript.

ACKNOWLEDGMENT

This work was supported by the NSF EFRI 2-DARE 1542879 and EAGER 1838412. P.M.D. and M.D. acknowledge support from NSF and NIH through the MRSEC DMR-1720530, NSF

EFRI 2- DARE (EFRI-1542707), NSF EAGER 1838456, and NIH R21 HG010536, as well as the Penn Grant for Faculty Undergraduate Research Mentorship (FURM) supporting undergraduate research. V.B.S acknowledges support from Army Research Office contract W911NF-16-1- 0447 and NSF grants EFMA-542879 and CMMI-1727717. We thank Dr. Robert Keyse at Lehigh University for assistance with STEM imaging and Dr. Bing Hao for assistance with the figures. The authors acknowledge the use of the Raman system supported by NSF Major Research Instrumentation Grant DMR-0923245. This work was carried out in part at the Singh Center for Nanotechnology, part of the National Nanotechnology Coordinated Infrastructure Program, which is supported by the National Science Foundation grant NNCI-1542153.

SUPPLEMENTARY FIGURES

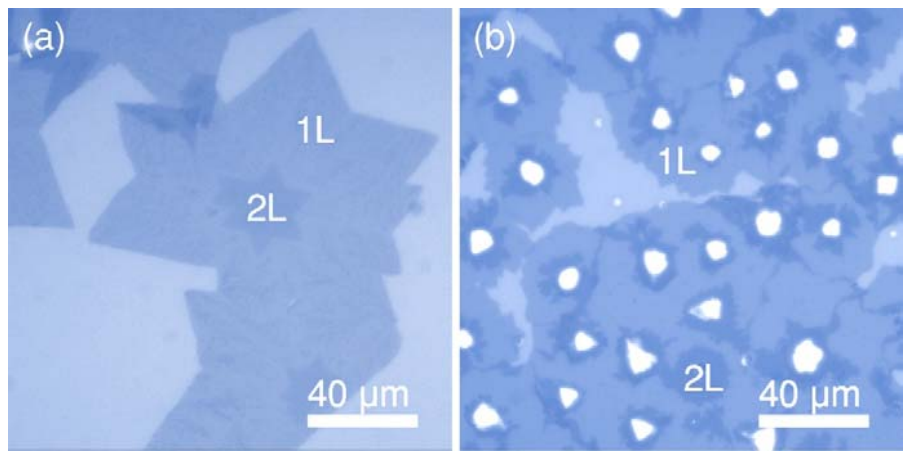

**Figure S1.** Optical images of the as-grown WSe$_2$ flakes using (a) 25 mM C$_{24}$H$_{40}$NaO$_5$ + 125 mM NaCl and (b) 25 mM C$_{24}$H$_{40}$NaO$_5$ + 500 mM NaCl as the growth promoter. The bright spots in (b) are multilayer WSe$_2$.

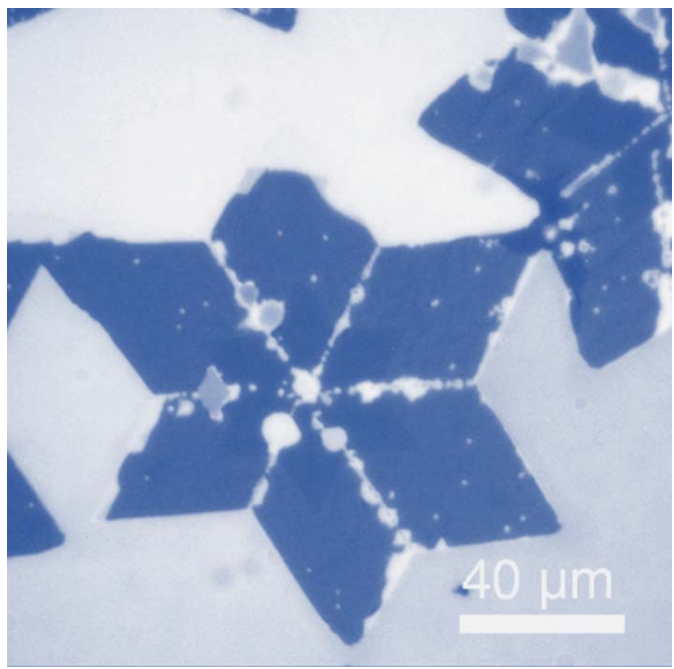

**Figure S2.** Optical image of a partially oxidized bilayer WSe$_2$ flake, in which the grain boundaries are clearly observed.

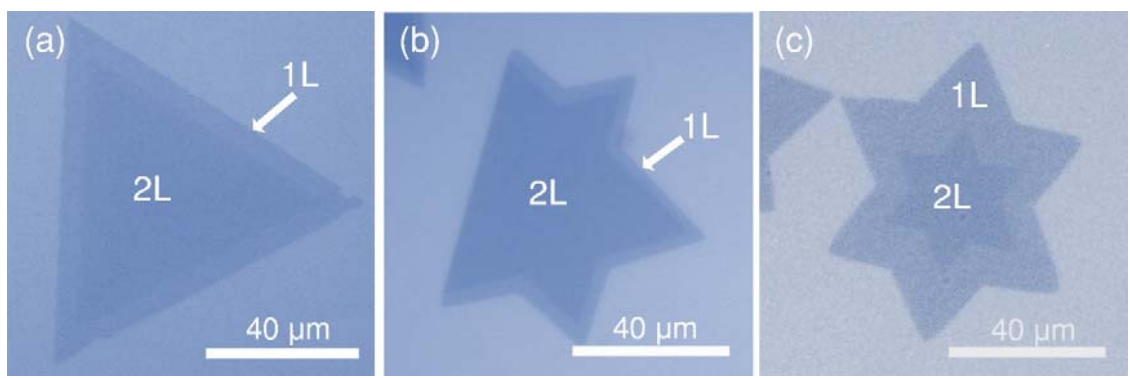

**Figure S3.** Optical images of bilayer WSe$_2$ flakes with varied morphologies using a mixture of 25 mM C$_{24}$H$_{40}$NaO$_5$ and 250 mM NaCl as the growth promoter.

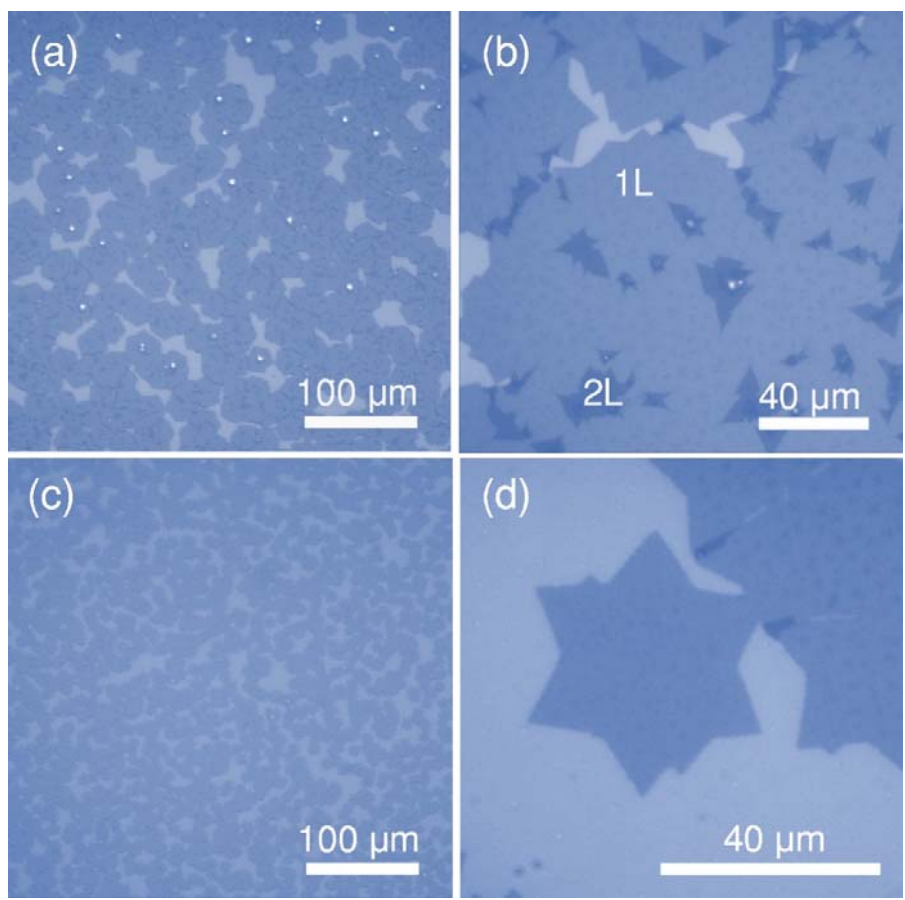

**Figure S4.** Optical images of the as-grown WSe$_2$ flakes using (a,b) 75 mM NaCl and (c,d) 25 mM NaCl as the growth promoter.

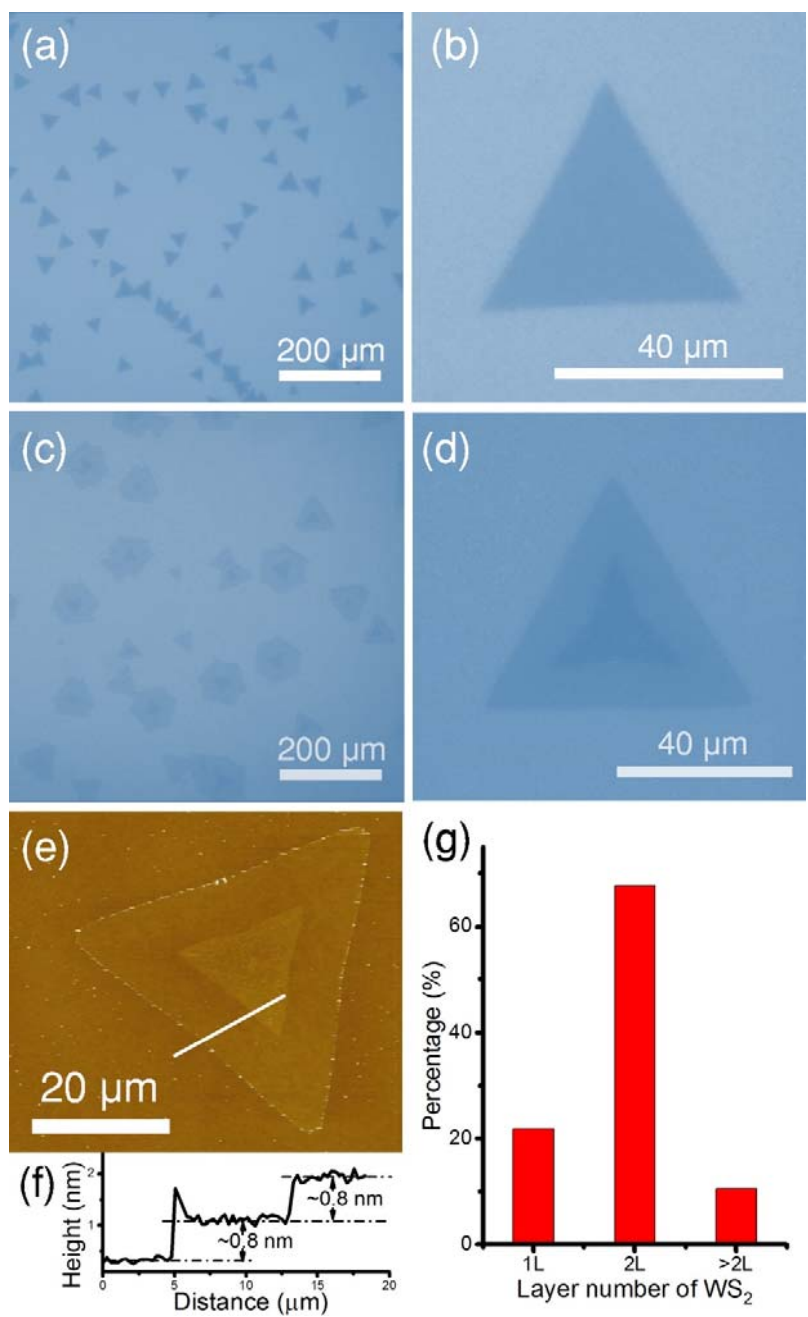

**Figure S5.** Optical images of WS$_2$ flakes grown by CVD using (a,b) 25 mM C$_{24}$H$_{40}$NaO$_5$ and (c,d) 25 mM C$_{24}$H$_{40}$NaO$_5$ and 250 mM NaCl as the growth promoter. (e) Atomic force microscope image of a bilayer WS$_2$ flake and (f) corresponding height profile; (g) Statistical analysis on the layer number of WS$_2$ flakes grown using 25 mM C$_{24}$H$_{40}$NaO$_5$ and 250 mM NaCl as the growth promoter, showing the preferential growth of bilayer material.

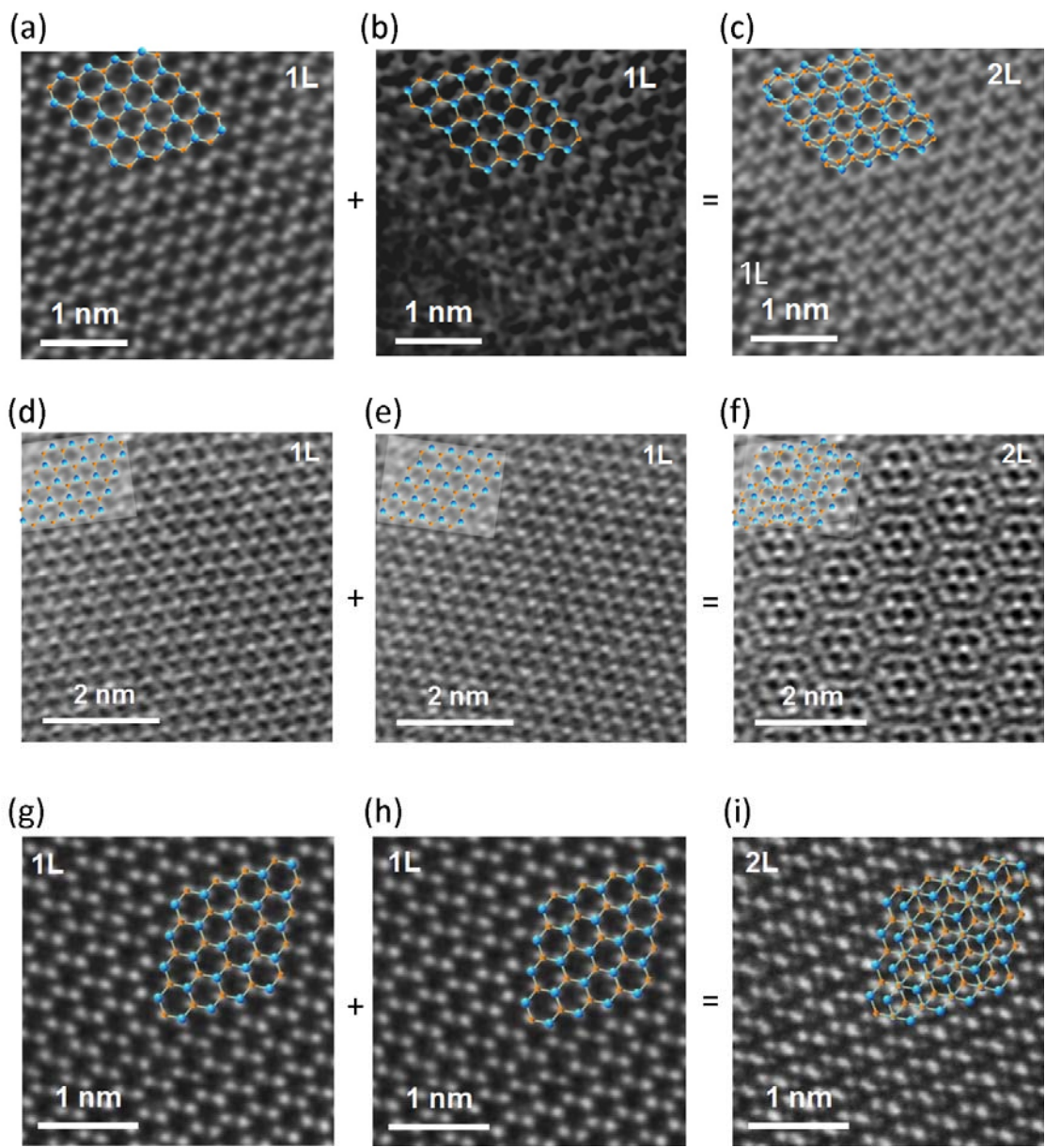

**Figure S6.** Analysis of the stacking orientation and atomic structure of bilayer WSe$_2$ by fast Fourier transform (FFT) decomposition of the 2L regions: (a-c) 60° twist angle or AA' stacking; (d-f) 15° twist angle; (g-i) 0° twist angle or AB stacking. In each set of the 3 images, the 2L regions is decomposed into two monolayers by FFT to identify the corresponding atomic models. The related inset images present the corresponding atomic models.

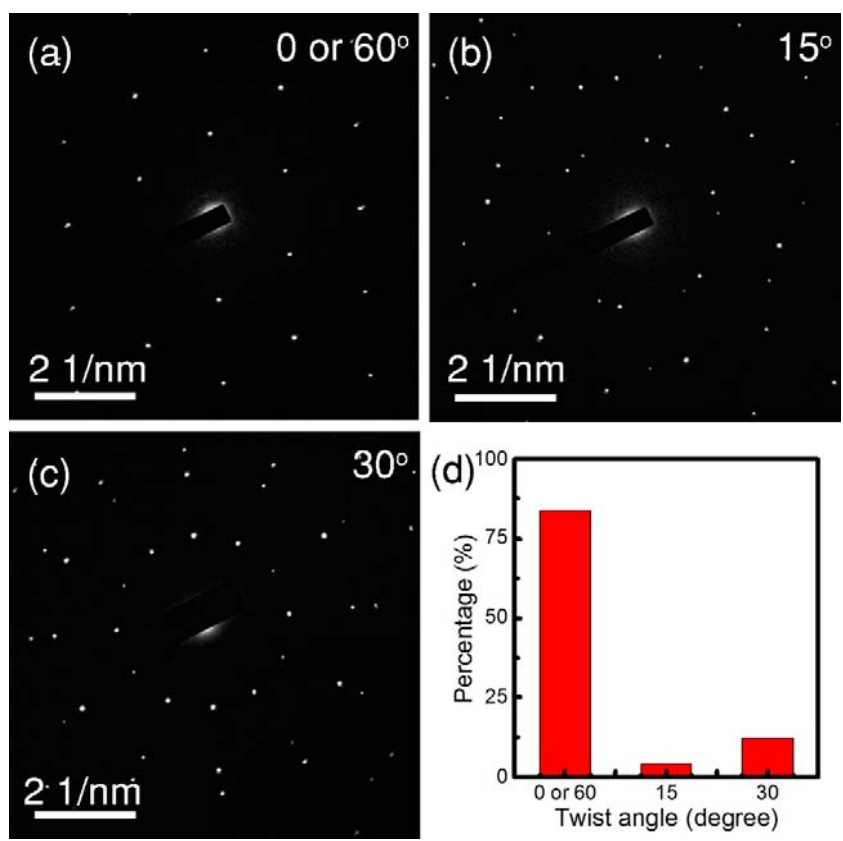

**Figure S7.** Selected area electron diffraction (SAED) patterns of bilayer WSe$_2$ flakes with (a) 0 or 60° twist, (b) 15° and (c) 30° twist angles; (d) Histogram of twist angle frequencies based on ~100 bilayer WSe$_2$ flakes.

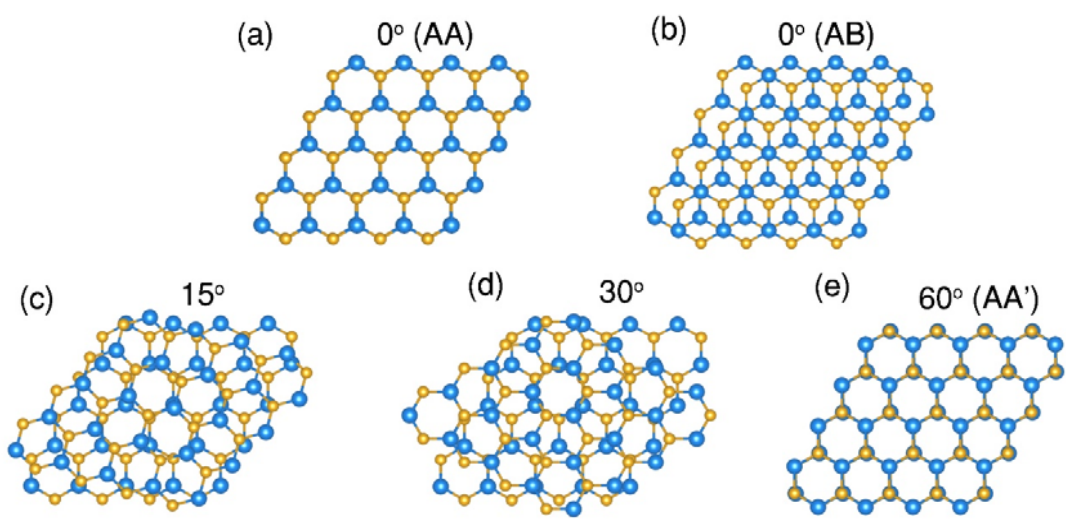

**Figure S8.** Summary of the atomic models of bilayer WSe$_2$ with varied stacking orientations.

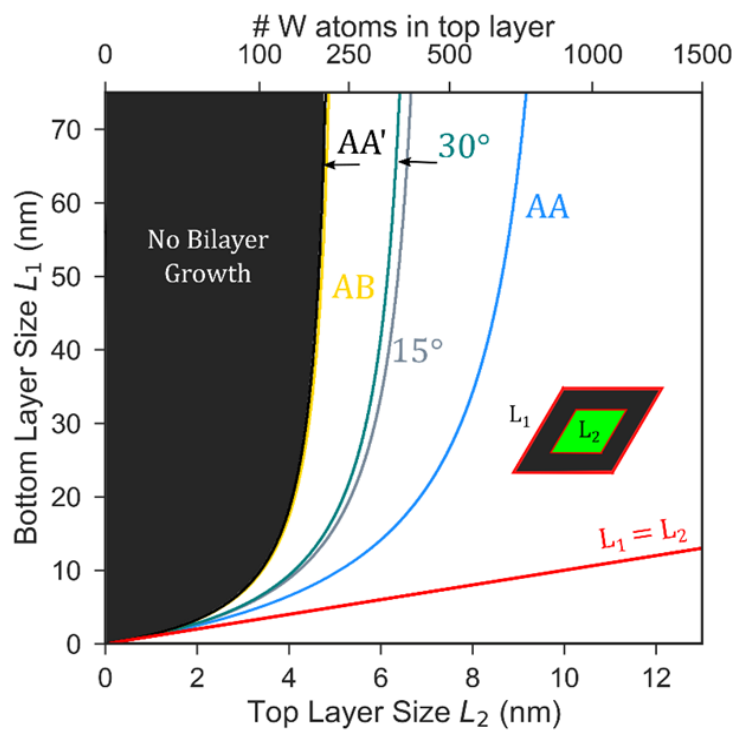

**Figure S9.** Thermodynamic growth diagram for bilayer $WSe_2$ flakes as a function of the instantaneous layer sizes L1 (bottom layer) and L2 (top layer). The lines mark the minimum required top layer size for thermodynamically favorable bilayer growth of triangular flakes over the range of bottom layer sizes.

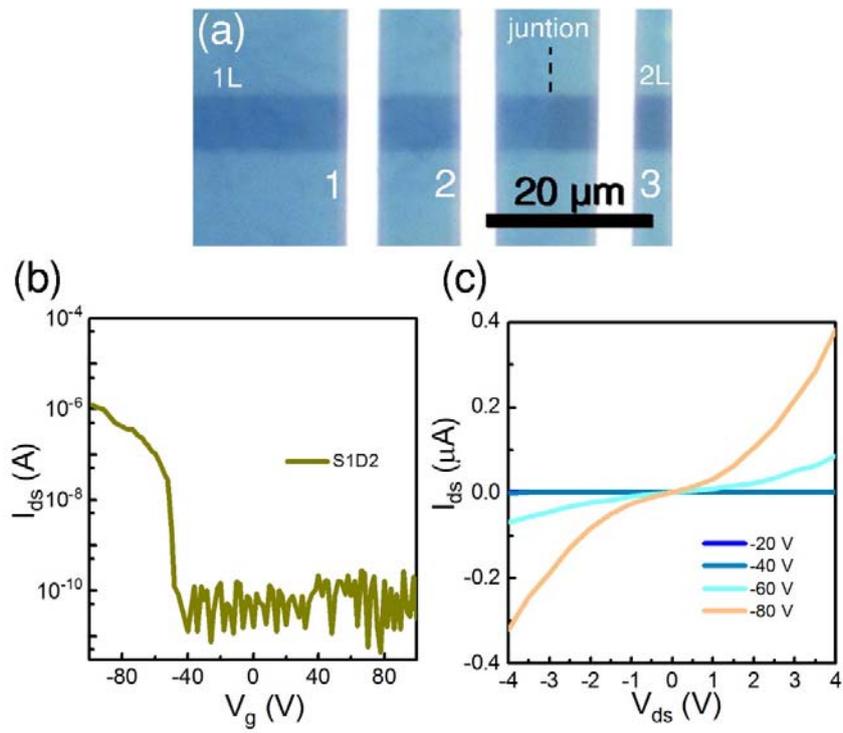

**Figure S10.** (a) Optical micrograph of a device structure based on a bilayer flake containing a large monolayer region. The electrode pair S1D2 contacts the monolayer region. (b) $I_{ds}$-$V_g$ curves for the S1D2 FET device. (c) $I_{ds}$-$V_{ds}$ curves for the WSe$_2$-based FET at different values of the backgate voltage.